
\input harvmac

\def\figin{\epsfcheck\figin}\def\figins{\epsfcheck\figins}
\def\epsfcheck{\ifx\epsfbox\UnDeFiNeD
\message{(NO epsf.tex, FIGURES WILL BE IGNORED)}
\gdef\figin##1{\vskip2in}\gdef\figins##1{\hskip.5in}
\else\message{(FIGURES WILL BE INCLUDED)}%
\gdef\figin##1{##1}\gdef\figins##1{##1}\fi}
\def\DefWarn#1{}
\def\figinsert{\goodbreak\midinsert}
\def\ifig#1#2#3{\DefWarn#1\xdef#1{fig.~\the\figno}
\writedef{#1\leftbracket fig.\noexpand~\the\figno}%
\figinsert\figin{\centerline{#3}}\medskip\centerline{\vbox{\baselineskip12pt
\centerline{\footnotefont{\bf Fig.~\the\figno:} #2}}}
\bigskip\endinsert\global\advance\figno by1}

\def\Title#1#2{\rightline{#1}\ifx\answ\bigans\nopagenumbers\pageno0
\vskip0.5in
\else\pageno1\vskip.5in\fi \centerline{\titlefont #2}\vskip .3in}

\font\caps=cmcsc10

\noblackbox
\parskip=1.5mm


\def\npb#1#2#3{{\it Nucl. Phys.} {\bf B#1} (#2) #3 }

\def\prd#1#2#3{{\it Phys. Rev. } {\bf D#1} (#2) #3 }
\def\prl#1#2#3{{\it Phys. Rev. Lett.} {\bf #1} (#2) #3 }

\def\ijmpa#1#2#3{{\it Int. J. Mod. Phys.} {\bf A#1} (#2) #3 }

\def\cmp#1#2#3{{\it Commun. Math. Phys.} {\bf #1} (#2) #3 }

\def\bb#1{{\tt hep-th/#1}}


\def\dj{\hbox{d\kern-0.347em \vrule width 0.3em height 1.252ex depth
-1.21ex \kern 0.051em}}



\lref\rcom{L. Susskind, L. Thorlacius and J. Uglum, \prd
{48}{1993}{3743.}}
\lref\rgib{G.W. Gibbons and M. Perry, {\it Proc. Roy. Soc. London}
{\bf A 358} (1978), 467.}
\lref\rgidd{S.B. Giddings, {\it ``Comments on Information Loss and
Remnants"}, Santa Barbara preprint UCSBTH-93-35, \bb{9310101.}}
\lref\rhow{S.W. Hawking, \cmp {43}{1975}{199.}}
\lref\rcall{C. Callan, S. Giddings, J. Harvey, and A. Strominger, \prd
{45}{1992}{1005.}}
\lref\rtho{G. t'Hooft, \npb {256} {1985}{727.}}
\lref\rthof{C.R. Stephens, G. t'Hooft and B.F. Whiting, {\it ``Black Hole
Evaporation without Information Loss"} Utrecht preprint THU-93/20,
hep-gc/ 9305008.}
\lref\rsvv{K. Schoutens, E. Verlinde and H. Verlinde, {\it ``Black Hole
Evaporation and Quantum Gravity".} CERN-Princeton preprint.
CERN-TH.7142/94, PUPT-1441 \bb{9401081.}}
\lref\rbom{L. Bombelli, R.K. Koul, J. Lee and R.D. Sorkin, \prd {34}
{1986} {373\semi}
M. Srednicki, \prl {71} {1993} {666\semi}
C. Callan and F. Wilczek, {\it ``On Geometric Entropy"}, preprint
IASSNS-HEP-93/87 \bb{9401072}\semi
D. Kabat and M.J. Strassler, {\it ``A comment on Entropy and Area"},
Rutgers preprint RU-94-10, \bb{9401125}\semi
J.S. Dowker, {\it ``Remarks on Geometric Entropy"}, Manchester
preprint, MUTP/94/2, \bb{9401159.}}
\lref\rsusk{L. Susskind, {\it ``Strings, black holes and Lorentz
contraction"}, Stanford preprint SU-ITP-93-21, \bb{9308139}\semi
L. Susskind, {\it ``Some Speculations about Black Hole Entropy in
String Theory"}, Rutgers preprint, RU-93-44 \bb{9309145}\semi
L. Susskind and J. Uglum, {\it ``Black Hole Entropy in Canonical Quantum
Gravity
and Superstring Theory"}, Stanford preprint. SU-ITP-94-1 \bb{9401070.}}
\lref\rbann{M. Ba{\~n}ados, D. Teitelboim and J. Zanelli, \prl
{69}{1992}{1849\semi}
G.T. Horowitz and D.L. Welch, \prl {71}{1993}{328.} }
\lref\rpol{J. Polchinski, \cmp {104}{1986}{37\semi}
M. Osorio \ijmpa {18} {1992}{4275.} }
\lref\real{E. Alvarez and M. Osorio, \prd {36}{1987}{1175.}}
\lref\rgro{D. Gross, M. Perry and L. Yaffe, \prd {25}{1982}{330.}}
\lref\rgrokleb{D. Gross and I. Klebanov, \npb {344}{1990}{475.}}
\lref\rkog{Ya. I. Kogan, JETP Lett. {\bf 45} (1987) 709\semi
B. Sathiapalan, \prd {35}{1987}{3277.}}
\lref\raw{J. Atick and E. Witten, \npb {310}{1988}{291.}}
\lref\rdeve{H.J. de Vega and N. Sanchez, \npb {299}{1988}{818.}}

\lref\rDVV{R. Dijkgraaf, E. Verlinde and H. Verlinde,
\npb {371} {1992} {269.}}



\line{\hfill PUPT-1448}
\line{\hfill {\tt hep-th/9402004}}
\vskip 1cm

\Title{\vbox{\baselineskip 12pt\hbox{}
 }}
{\vbox {\centerline{Horizon divergences}
\medskip
\centerline{of Fields and Strings}
\medskip
\centerline{in Black Hole backgrounds}  }}

\centerline{$\quad$ {\caps J. L. F. Barbon}}
\smallskip
\centerline{{\sl Joseph Henry Laboratories}}
\centerline{{\sl Princeton University}}
\centerline{{\sl Princeton, NJ 08544, U.S.A.}}
\centerline{{\tt barbon@puhep1.princeton.edu}}
\vskip 0.4in

 General  arguments based on curved space-time thermodynamics show
that any extensive quantity, like the free energy or the entropy of
thermal matter, always has a divergent boundary contribution in the
presence of event horizons, and this boundary term comes with the
Hawking-Bekenstein form. Although the coefficients depend on the
particular geometry we show that intensive quantities, like the
free energy density, are universal in the vicinity of the horizon.
 From the point of view of the matter degrees of freedom this
divergence is of infrared type rather than ultraviolet, and we
use this remark to speculate about the fate of these pathologies
in String Theory. Finally we interpret them as instabilities of
the Canonical Ensemble with respect to gravitational collapse
via the Jeans mechanism.


\Date{1/94}


\newsec{Introduction}

The behaviour of quantum fields in black-hole backgrounds is plagued
with paradoxes that are commonly viewed as important keys towards
a full understanding of Quantum Gravity. They usually hit on
non-perturbative or non-linear aspects of gravity which are far beyond
the perturbative answers, and  very difficult to tackle in short
distance modifications like String Theory in its present formulation.

 The most famous of such troubles is
probably the Hawking information paradox \refs\rhow , according to which it
seems impossible to define a unitary $S$-matrix for scattering processes
in the presence of event horizons. This problem has received renewed
attention in recent years, specially through the study of
two-dimensional toy models \refs\rcall . In spite of the
interesting work done, the
whole field is still divided in ``schools of thought" regarding this
problem, and it seems increasingly evident that some very subtle
mechanism is behind this paradox.

 In this paper we examine the analytic continuation to imaginary
time of the same problem. It was pointed out by 't Hooft \refs\rtho\
 that the
naive canonical ensemble free energy of free fields in the exterior
of the Schwarzschild black hole diverges linearly as the radial
coordinate  approaches the horizon. So, in this case
the trouble in defining a reasonable evolution operator appears
as a divergence of the trace of the canonical density matrix.
In fact, the paradox in    the euclidean version is even sharper.
While we could get used to live with non-unitary evolution, it
seems impossible to accept a divergence in a physical quantity.
In this sense, the analogy with the ultraviolet catastrophe of the
black body radiation is almost complete.

 These horizon divergences are very general, since they follow
from simple classical thermodynamical arguments and the Equivalence
Principle. Let us consider a
system in thermal equilibrium at asymptotic temperature $T$ in a
black-hole background of the form
\eqn\metric
{ds^2 =-g_{00} dt^2 + h_{\alpha\beta} dx^\alpha dx^\beta = - \lambda (r) dt^2 +
{1\over \mu(r)} dr^2 + r^2 d\Omega_{d-2}}

The functions $\lambda(r)$ and $\mu(r)$ are such that the metric goes
to Minkowski space at $r=\infty$ and we have a non-degenerate horizon
at finite radial coordinate $r_0$ i.e. $\lambda(r) \sim \lambda'_{0}
(r-r_0) $ , $ \mu(r) \sim \mu'_{0} (r-r_0) $ with associated Hawking
temperature $4\pi T_H = \sqrt{\lambda'_{0} \mu'_{0}}$ .
 According to the Equivalence Principle for a system in thermal
equilibrium the local temperature times $\sqrt{-g_{00}}$ is constant,
and equal to the asymptotic temperature $T$. So, for the
calculation of any extensive dimensionless quantity like the entropy
or $\beta F$ we can divide the system in small boxes  such that the
metric is locally  constant and simply red-shift the
local temperature dependences of the flat space quantities. For large
enough $T$ this gives,
  \eqn\exten
{\beta F \sim \int d^{d-1} x \sqrt{h} \left({T\over
\sqrt{-g_{00}}}\right)^{d-1}}
and we see the origin of the divergence in the standard red-shift of
the temperature. Even if we arrange for a small local curvature at
the horizon the consideration of canonical equilibrium at any finite
$T$ leads to infinite energy densities at the horizon. It is useful to
rewrite \exten\ in the form
\eqn\optical
{\beta F \sim T^{d-1} {\overline V}_{d-1}}
where $\overline V$ is the so-called optical volume, computed out
of the optical metric defined as ${\overline g}_{\mu\nu} = g_{\mu\nu}
/ (-g_{00}) $ \refs\rgib . This conformally related metric
     is useful because it
allows us to talk about divergences of the optical volume instead of
local temperatures. In particular, from \optical\ we expect the
boundary infinities to depend mainly on geometrical details of the
manifold near the horizon. Indeed, expanding the metric around $r_{0}$
we have the estimate,
\eqn\bek
{\beta F \sim S \sim {A_{H}^{(d-2)} \over \epsilon^{d-2}} \left({T\over
T_H}\right)^{d-1}}
where $A_H$ denotes the horizon area. We see that we recover quite
generally the Hawking-Bekenstein form with an explicit cutoff in
proper distance $\epsilon$, usually taken of the order of the Planck
length (in the two-dimensional
case we get a logarithmic divergence). This formula is also true for a
thermal ensemble of gravitons, and \bek\ can be viewed as a quantum
correction to the Hawking-Gibbons classical term, following from the
leading instanton contribution to the euclidean path integral in the
gravitational sector.
 These  boundary terms have the same form as the so-called geometric
entropy that results from tracing over degrees of freedom inside
certain boundary \refs\rbom . On the other hand, there is
a growing feeling that one should resist temptation of talking about
the "inside" when asking "outside" questions
   \refs{\rcom,\rsvv,\rthof} .
 For this reason we will
try to discuss these divergences in terms of exterior data only.

 In the remainder of this paper we shall state \exten\ in a more
detailed way. In the next section we show that a WKB approximation
to the canonical ensemble partition sum allows us to calculate the
free energy density, which becomes equal to the red-shifted flat
space result when we approach the horizon. This confirms that the
boundary infinity is governed only by the divergence of  the optical
volume. From the point of view of the spectrum of the matter theory,
the partition sum divergence looks of infrared type, and this allows
us to speculate that the one loop string free energy is not safe from
this disease. This is interesting in relation to the program outlined
by Susskind et al. \refs\rsusk , according to which perturbative String
Theory would play an important role in the resolution of the horizon
problems.
 Finally, we interpret these divergences as yet another (although more
drastic) instability of the Canonical Ensemble with respect to
gravitational collapse.

 \newsec{WKB estimates}

 In this section we prove \exten\ for the case of a scalar field in the
exterior of the black hole \metric\ in the WKB approximation.
 Our aim is to compute the thermal free energy density as
\eqn\canon
{F = -{1\over\beta} {\rm log} {\rm Tr} e^{-\beta H} = \int d^{d-1} x
\sqrt{h} f (\beta, x)}
To one loop, after standard Bose oscilator algebra we have:
\eqn\loop
{F_{1loop} = {1\over\beta} \sum_i {\rm log} (1- e^{-\beta \omega_i})}
The sum runs over single particle states of $H = \sum_i
\omega_i a_{i}^{\dagger} a_{i} $, the usual normal ordered Hamiltonian
determined by the frequency modes with respect to the
Schwarzschild-like coordinate $t$,
\eqn\frec
{\partial_t \phi_{\omega} = -i\omega\phi_{\omega}}
where $\phi_{\omega}$ solves the generalized Klein-Gordon eq in the
space-time \metric
\eqn\KGeq
{(\nabla_{x}^{2} - m^2 - \xi R(x))\phi_{\omega}(x) = 0}
In solving for the energy spectrum it proves convenient to decompose
$\phi$ according to the symmetries in the form
\eqn\harm
{\phi_{n,\alpha} (r) = {e^{-i\omega_{n,\alpha}
t}\over\sqrt{2\omega_{n,\alpha}}} r^{2-d\over 2} f_{n,\alpha}(r)
Y_{\alpha} (\Omega)}
Here $n$ is a radial quantum number and $\alpha$ summarizes the
angular quantum numbers. $Y_{\alpha} (\Omega)$ is a normalized
eigenfunction of the angular Laplacian (spherical harmonic in $d=4$
and integer exponential in $d=3$). If we further change the radial
variable to
\eqn\tortoise
{x = \int^{r} \sqrt{g_{rr}\over -g_{00}} \equiv \int^{r} {1\over
\gamma (r)}}
then the Klein-Gordon inner product has the form
\eqn\kleinGo
{\eqalign{
\langle \phi_{n,\alpha} \mid \phi_{n'{\alpha}'}\rangle_{KG} &= i
\int_{t={\rm const}} d\Sigma^{\mu}
\phi_{n,\alpha}^{\ast}
\mathrel{\mathop \partial^\leftrightarrow}_\mu
\phi_{n',{\alpha}'} \cr
&= {\omega + {\omega}' \over 2\sqrt{\omega{\omega}'}}
e^{i(\omega-\omega')t} \delta_{\alpha, \alpha'} \int_{\cal D} dx
f_{n, \alpha}^{\ast} (x) f_{n',\alpha'} (x)} }
where $\cal D$ is the image of the exterior radial domain $(r_0 ,
+\infty)$ under the mapping \tortoise\ to ``tortoise-like" coordinates.
So the Klein-Gordon inner product is normalized once the $f's$ are
$L^2$ normalized on $\cal D$. This suggests that, under this change of
variables, the equation \KGeq\
becomes a Schr{\"o}dinger problem.
\eqn\schr
{\left\{ -{1\over 2}{d^2 \over dx^2} + V_{\alpha} (x) \right\}
f_{n,\alpha} (x) = {\omega_{n,\alpha}^{2} \over 2} f_{n,\alpha} (x) }
where $V_{\alpha}$ is the effective potential:
\eqn\pot
{V_{\alpha}(x) = {d-2 \over 4} \gamma(r)\left[{\gamma' (r)\over r} +
{d-4 \over 2r^2} \gamma (r) \right] + {\lambda (r)\over 2}
\left[{\Delta_{\alpha} \over r^2} + m^2 + \xi R(r)\right] }
Here prime denotes $ d/dr$ and the $x$-dependence is implicit via \tortoise.
$\Delta_{\alpha}$ is the eigenvalue of the angular Laplacian
($l(l+1)$ for $d=4$ and $l^2$ for $d=3$).
 It turns out that only certain universal features of this eigenvalue
problem are important for us. First of all, the new domain $\cal D$ is
non-compact and infinite, $x\in (-\infty, +\infty)$ and
asymptotically $x \sim r$ as $r\rightarrow \infty$ but behaves
logarithmically near the horizon,
\eqn\strech
{r-r_0 \sim e^{4\pi T_H x} }
so that, in these coordinates, the horizon appears as an open boundary.
Second, the asymptotics of the potential are,
\eqn\potasym
{V_\alpha (x) \sim {m^2 \over 2} \,\,\,\,\, {\rm as}\,\,\,\,\, r\rightarrow
\infty}
while near the horizon it decreases exponentially,
\eqn\tail
{V_\alpha (x) \sim C e^{4\pi T_H x}  \,\,\,\,\, {\rm as}\,\,\,\,\, r\rightarrow
r_0 }
where the constant $C$ depends on details of the field and the
background. Since we are  interested only in the positive definite
spectrum,  according to \schr, we obtain continuous spectrum for the
field even if we cutoff at large volume putting the system inside a
box. If we regulate the non-compact boundary at the horizon by
imposing some Dirichlet condition at $x_{-}$ ('t Hooft's brick-wall
model), the lowest positive
eigenvalue goes to zero as we remove this cutoff independently of the
mass of our field. Thus from the point of view of the field energy
levels $\omega_{n,\alpha} $ the problem seems of infrared nature.

Since in both asymptotic regions the potential looks flat, we expect
the WKB approximation to give accurate results. To compute the free
energy density we start from the extensive quantity,
$$
F = {1\over\beta} \sum_{\alpha} \sum_{n(\alpha)} {\rm
log}(1-e^{-\beta\omega_{\alpha,n}})
$$
After replacing the radial sum by integrals and integrating by
parts we have,
$$
F = -\sum_{\alpha} \int d\omega {n_\alpha \over e^{\beta\omega} - 1}
$$
with $n_\alpha$ given by the WKB ansatz:
$$
n_{\alpha} (\omega) \sim {1\over \pi} \int_{x_{-}}^{x_{+}}
 dx \sqrt{\omega^2 - 2V_{\alpha} (x)}
$$
and the  integrals run over the values that keep the argument
of the square root positive (under this approximation one can see that
the boundary terms in the previous manipulations vanish).
   Now, since for $x_{-} \ll 0$, $V_{\alpha}
(x_{-})$ is almost independent of  $\alpha$, we may permute the angular
sum with the integrals to get,
$$
F \sim \int_{\sqrt{2V_s (x_{-})}}^{\infty} {d\omega \over
e^{\beta\omega} - 1} \int_{x_{-}}^{x_{+}} dx \sum_{\alpha}
\sqrt{\omega^2 - 2 V_{\alpha} (x)}
$$
where $V_s (x)$ denotes the $s$ wave potential ($\Delta_{\alpha} = 0$).
Next we approximate again the $\alpha$ sum by an integral:
$$
\sum_{\alpha} \sqrt{\omega^2 - 2V_{\alpha}(x)} \cong C_d \left( {r\over
\sqrt{-g_{00}}}\right)^{d-2} \left(\int d\Omega_{d-2}\right)
\left(\sqrt{\omega^2 - 2 V_s (x)}\right)^{d-1}
$$
For example, $C_4 = 1/4\pi$ and $C_3 = 1/4$.
Finally, permuting $x$ and $\omega$ integrals and transforming back to
standard exterior coordinates we recognize the free energy density
given by the expression,
\eqn\freeen
{f_{WKB} (r,\beta) =-{C_d \over \pi} {\beta^{-d} \over (\sqrt{-g_{00}})^{d-1}}
\int_{\beta\sqrt{2V_s (r)}}^{\infty} {dz (z^2 - 2\beta^2 V_s (r))^{d-1
\over 2} \over e^z - 1} }

For the ``bulk" contribution, far away from the horizon,
we obtain the standard result (with the correct numerical factors),
$$
f(r\rightarrow \infty) \cong -{C_d \over \pi} T^d  \int_{\beta
m}^{\infty} {dz (z^2 - (\beta m)^2)^{d-1 \over 2} \over e^z - 1}
$$
and close to the horizon we get a universal expression in the sense
that it is independent of the curvature coupling $\xi$ or the mass,
\eqn\frehor
{f(r\rightarrow r_0) \cong -{C_d \over \pi} T \left( {T\over
\sqrt{-g_{00}}}\right)^{d-1} \Gamma(d) \zeta(d) }
 This formula can be considered as a check of \exten. The interesting
lesson for us is that, when propagating very close to the horizon all
fields behave as massless, and possible curvature couplings become
irrelevant (for fermions only some numerical factors change, and
higher spin or extra quantum numbers like flavor, etc amount simply to
an integer prefactor).
 So, technically, the disease seems to be infrared and this puts
into
question the regularity of strings under these circumstances. These
present a very good ultraviolet behaviour, but their infrared
problems are at least as severe as in Field Theory.

 There are a few minor changes in this analysis for two interesting
particular cases. Namely, Rindler space and the 3d black-hole of
Ba{\~n}ados et. al. \refs\rbann . Both spaces fail to approach
Minkowski space    at large $r$.
In the first case the effective potential grows exponentially at
infinity and there is no need for external infrared cutoff (external
box). The 3d black-hole approaches anti-De Sitter at long distances
and the effective potential grows quadratically with $r$ unless we consider
the tachyon, for which one gets an unstable potential at long
distances. However, no modifications arise in these examples for the
horizon region.

\newsec{Euclidean Formalism}

Before discussing Strings it is useful to obtain first quantized path
integral representations for the Field Theory case. These are defined
starting from the euclidean formalism.

 Naively we would write the total free energy as
\eqn\naive
{F(\beta) = -{1\over\beta} {\rm log} \int {\cal D} \phi e^{-S_\beta} =
{1\over 2\beta} {\rm log}{\rm det} (-\nabla_{E}^{2} +m^2 + \xi
R) }
The euclidean action
\eqn\eucliac
{S_\beta (\phi) = {1\over 2} \int_{0}^{\beta} d\theta \int \sqrt{g} \phi
(-\nabla_{E}^{2} +m^2 +\xi R)\phi }
is defined on the euclidean section of \metric\
\eqn\euclmetric
{ds_{E}^{2} = g_{00} d\theta^2 + h_{\alpha\beta} dx^{\alpha}dx^{\beta}
}
with periodic identification $\theta \equiv \theta + \beta$.
Ultraviolet divergences in \naive\ amount to mass renormalization or
normal ordering (subtraction of the zero point energy). However
\naive\ leads to a paradox because the eigenvalue problem of the
scalar Laplacian in the euclidean manifold \euclmetric\ is perfectly
well defined once we impose a radial cutoff at long distances. It
seems that there is no room for the infrared divergences discussed in
the previous section. This is specially clear for the particular case
$\beta = 1/T_H$, where even the conical singularity at $r=r_0$ is
absent. For example, we may think of the 2d euclidean black hole (the
cigar metric). If we cutoff at large $r$ we have a smooth compact
manifold with disc topology in which the Laplacian has discrete
eigenvalues without any further Dirichlet condition imposed. Hence we
must conclude  that our divergent free energies of the previous
section are not given by \naive.

 In fact, the resolution of this paradox is both easy and instructive.
Let us start  from \loop\ and apply some well known algebraic
tricks. For example, we can write
\eqn\zetaf
{F(\beta) = -{1\over\beta}{\rm log}\prod_{n,\alpha}
(1-e^{-\beta\omega_{n,\alpha}}) = {1\over 2\beta} {\rm log}
\prod_{m,n,\alpha} \left({4\pi^2 m^2 \over \beta^2} +
\omega_{n,\alpha}^{2}\right) - {1\over 2} \sum_{n,\alpha}
\omega_{n,\alpha} }
The right hand side of this equation is understood in the sense of
zeta function regularization. It has "log det" form after subtraction
of the zero point energy (cosmological constant $\Lambda_B$). So,
instead of \naive\ we have
\eqn\good
{F(\beta) = -{1\over \beta} {\rm log}{\rm det}^{-1/2} [ g_{00}
(-\nabla_{E}^{2} +m^2 + \xi R)] - \Lambda_B }
where the differential operator has eigenfunctions,
\eqn\eucleigen
{\phi_{m,n,\alpha} = {e^{2\pi i m\theta /\beta}\over \sqrt{\beta}}
r^{2-d \over 2} f_{n,\alpha} (r) Y_{\alpha} (\Omega) }
The difference between \naive\ and \good\ is simply the $g_{00}$
factor in front ensuring that the "time" component of the eigenvalues
is of the form $4\pi^2 m^2 /\beta^2$. As for the Field Theory path
integral representation, the euclidean action remains the same, but
the measure is regularized with respect to the inner product
\eqn\kgtres
{\langle \phi_1 \mid \phi_2 \rangle = \int d\theta d^3 x \sqrt{h\over
g_{00}} \phi_{1}^{\ast} (\theta, {\vec x}) \phi_2 (\theta,{\vec x})}
This leads to the formal measure
$$
{\cal D}\phi \rightarrow \prod_x {d\phi(x) \over
\sqrt{2\pi}}\left({h\over g_{00}}\right)^{1/4} (x)
$$
Now we can see the origin of the horizon singularities. The euclidean
manifold is compact but the inner product \kgtres\ is ill-defined at
the horizon. The modes must oscillate wildly at $r\sim r_0$ if we want
\kgtres\ to be finite. In fact, the spatial section is precisely the
Klein-Gordon inner product previously introduced and that, in turn,
explains the form of the eigenfunctions in \eucleigen. On the other
hand, in \naive\ the determinant is defined with respect to the
generally covariant inner product, with the standard volume element
$d\theta d^3 x \sqrt{g}$, which is perfectly regular at $r=r_0$.

 The first-quantized path integral follows most easily from the
expression \zetaf.  Using the integral representation of the logarithm:
\eqn\param
{F(\beta) = -\Lambda_B -{1\over 2\beta} \int_{0}^{\infty} {ds\over s}
\sum_{m,n,\alpha} e^{-s({4\pi^2 m^2 \over \beta^2}
+\omega_{n,\alpha}^{2} )} }
and recalling \schr\ we have
\eqn\preint
{F(\beta) = -\Lambda_B - {1\over 2\beta} \int_{0}^{\infty} {ds\over s}
{\rm Tr} e^{ -s\left( -{d^2 \over d\theta^2} - {d^2 \over dx^2}
-{g_{00} \over r^2} \nabla_{\alpha}^{2} + 2V_s (x)\right) }}
where $\nabla_{\alpha}^{2}$ stands for the angular Laplacian and $V_s
(x)$ is the $s$-wave effective potential.

 Now, going to a path integral representation and formally changing
variables to $(r,\Omega)$ we get,
\eqn\pathint
{F(\beta) = -\Lambda_B -{1\over 2\beta} \int_{0}^{\infty} {ds\over s}
\oint {\cal D}x(\tau) e^{-S[x(\tau)]} }
where the world-line action is
\eqn\wlaction
{S[x(\tau)] = {1\over 4} \int_{0}^{s} d\tau {\bar g}_{\mu\nu}
{dx^{\mu} \over d\tau} {dx^{\nu}\over d\tau} + 2\int_{0}^{s} d\tau V_s
(r(\tau)) }
and the measure ${\cal D}x(\tau) \sim \prod_{\tau,\mu} dx^{\mu} (\tau)
\sqrt{{\bar g}(x(\tau))}$. Here $ {\bar g}_{\mu\nu} =
g_{\mu\nu}/g_{00} $ is the (euclidean) optical metric defined in the
introduction,	 and we  recover the rule heuristically stated in
\optical. The calculation of extensive quantities in the Canonical
Ensemble makes use of the optical metric as effective geometry.

Note that, since ${\bar g}_{00} =1$ the $\beta$-dependent part of the
path integral is easy to compute by the standard soliton
decomposition: $\theta = m\beta\tau /s + \theta '(\tau)$ with $\oint
d\theta ' = 0$. The result agrees with \param\ after Poisson
resummation,
$$
\beta \sum_m e^{-{\beta^2 m^2 \over 4s}} = 2\sqrt{\pi s} \sum_m
e^{-{4\pi^2 m^2 s \over \beta^2}}
$$
Incidentally, dropping the $m=0$ term in the left hand side amounts to
the vacuum energy subtraction, and we can write the expression,
\eqn\stranal
{F(\beta) = \int_{0}^{\infty} {ds\over s^2} \sum_{m\neq 0} e^{-m^2
\beta^2 \over 2\pi s} \Lambda (s) }
where
$$
\Lambda (s) = -{1\over 2} \sqrt{s\over 2} \sum_{n,\alpha} e^{-{\pi\over
2} s \omega_{n,\alpha}^2}
$$
This form is useful in discussing the relation between strings and
point particles at this level.

 Going back to \param\ we see that vanishing eigenvalues
$\omega_{n\alpha}$ give rise to divergences at large proper times,
formally similar to the $\tau_2 \rightarrow \infty$ infrared
divergences one encounters in String Theories.

\newsec{Strings}

 The string analog of \pathint\ would be some modular integral of the
form
\eqn\strpath
{F(\beta) = -\Lambda_B + \int_{\cal F} {d^2 \tau \over \tau_{2}^{2}}
Z_{gh} (\tau, {\bar \tau}) \int {\cal D} X e^{- S(X,\beta)} }
where ${\cal F}$ denotes the genus one fundamental modular domain:
$\tau = \tau_1 + i \tau_2$, $-1/2 \leq \tau_1 \leq 1/2$, $\tau_2 \geq
0$, $|\tau | > 1$. $Z_{gh}$ is the ghost partition function and the
sigma-model partition function computes the sum over all embeddings of
the flat torus onto the optical target space with action:
\eqn\wsaction
{S(X) = {1\over 2\pi} \int_T \left( {\bar g}_{\mu\nu} \partial
X^{\mu} {\bar \partial}X^{\nu} + T(X) + \Phi(X) R^{(2)} + ...\right)}
Here ${\bar g}_{\mu\nu}$ is the optical metric and we have included a
possible dilaton term and a tachyon background such that the field
equations of motion reproduce the potential $V_s (r)$.

It is clear that the presence of the tachyon background renders the
previous sigma-model path integral as a purely formal expression.
Tachyon backgrounds are very difficult to handle even at leading order
in a low energy expansion ($\alpha '$ expansion). However, at least for
the one loop free energy there is a physical recipe that gives the
right answer in all those cases that the path integral is computable.
Since the one loop thermal partition function reduces to counting
states with the tree level spectrum as weights, we hope that the
thermal free energy of a string ensemble equals the sum of the free
energies for the fields in the string spectrum. This is technically
termed as the "analog model" for finite temperature computations, and
it is known to give the same answer as the path integral bellow the
Hagedorn temperature \refs\rpol . (Above $T_{Hag}$ the
partition sum itself does not  converge.)

 In fact, this result sounds strange at first  sight because there are
at least two features of the string  one loop path integral that seem
specifically  "stringy". Namely, the integration region in moduli
space avoids the ultraviolet $\tau_2 \rightarrow 0$ region by a sort
of modular invariant cutoff procedure, and further the string has
extra "winding" states without point particle counterpart.
Interestingly enough, it turns out that these two features cancel each
other provided the integrand over moduli space is modular invariant.
In our case \strpath\ it is precisely the use of the optical sigma
model what allows us to prove such cancellation, due to the fact that
the "temporal" partition function factorizes as a free scalar field
(${\bar g}_{00} = 1$):
$$
S(X) = {1\over 2\pi} \int \partial \theta {\bar \partial} \theta +
S_{d-1} (r,\Omega)
$$
As in the point particle case, the $\beta-$dependent part of the
integrand over the moduli space equals the flat space result to all
orders in perturbation theory. For  genus one  we have,
$$
Z_{\theta} (\beta,\tau) = \beta\,\, \Theta (\beta,\tau) {\tau_{2}^{-1/2}
\over \eta {\bar \eta}}
$$
where $\eta (\tau) = e^{i\pi\tau /2} \sum_{n=1}^{\infty} (1-e^{2\pi i
\tau n})$ is the Dedekind eta function and $\Theta$ is the  Riemann
theta function coming from the soliton sum,
$$
\Theta (\beta,\tau) = \sum_{n,m =-\infty}^{+\infty} e^{-{\beta^2 \over
2\pi \tau_2} |m\tau + n|^2}
$$
 Here $m$ and $n$ are winding numbers of the torus homology cycles:
$$
\oint_a d\theta = n\beta \,\,\,\,\,\,\, \oint_b d\theta = m\beta
$$
As in the point-particle case we can subtract the vacuum energy (here
U.V. finite) by constraining $m,n \not= 0$ in the winding sums, and we
get,
\eqn\modinv
{F(\beta) = \int_{\cal F} {d^2\tau \over \tau_{2}^{2}} \left[
\Theta(\beta, \tau) -1\right] {\bar \Lambda}(
\tau, {\bar \tau}) }
where ${\bar \Lambda} (\tau, {\bar \tau})$ is the integrand of the
cosmological constant for the optical sigma model (including the ghost
piece). Now, if we assume that $\bar \Lambda$ is modular invariant we
can extend the modular integration over $\cal F$ to the strip $\cal
S$: $\tau_2 > 0$, by extending the integrand by cosets of the
translation subgroup $\tau \rightarrow \tau +1$, using the remarkable
identity ($\tau = \theta +i s$),
\eqn\truco
{\int_{0}^{\infty} {ds \over s^2} \int_{-1/2}^{1/2} d\theta
\sum_{m\not= 0} e^{-{\beta^2 m^2 \over 2\pi s}} = \int_{\cal F} {d^2
\tau \over \tau_{2}^{2}} \sum_{m,n \not= 0} e^{-{\beta^2 \over 2\pi s}
|n\tau + m|^2}}
Now,  on the left-hand side we have a "proper time" integral of the
same form as in \stranal. The $\theta$ integral is spectator here,
but it enforces the level matching on the right hand side.
Symbolically,
$$
\int_{-1/2}^{1/2} d\tau_1 \sim \delta (L_0 - {\bar L}_0)
$$
and we see that indeed the winding modes act as a modular invariant
regulator of the U.V. region. For well defined models these
manipulations require good behaviour of $\Lambda (\tau)$, which
basically means absence of infrared divergences as $\tau_2
\rightarrow \infty$. So, one is led in general to supersymmetric
strings, for which  a number of complications in handling the spin
structures arise, but the general idea of trading the sum over
windings by the U.V. region in moduli space remains true. One should
note that these tricks require in general $\beta \not= 0$. For
example,
modular invariant extensions of the cosmological constant as
calculated in the analog model are very formal, since the latter is
always U.V. divergent in field theory.

 Of course we do not have a Conformal Field Theory expression for
${\bar \Lambda}(\tau,{\bar \tau})$ in the optical background, but the
fact that we were able to prove the first step towards an analog
model version is encouraging. Had we started from the path integral in
the physical metric  the $\beta-$dependence would not be explicit.
Indeed, from the point of view of Statistical Mechanics the analog
model provides the definition based on "counting" and hence it can be
regarded as more fundamental than the path integral one. So in our
case we may take the analog model as a definition and consider the
previous comments as heuristic evidence  that a path integral
representation makes use of the optical background.

 An explicit construction of ${\bar \Lambda}(\tau,{\bar \tau})$ along
the lines of the analog model for some particular cases might not be
completely out of reach. Given the string spectrum for some exact
background one should start from the sum of the field free energies
and try to apply \truco\ backwards. There are two cases where the
basic ingredients are known. One is the two-dimensional black-hole
where we have basically one massless field and we may use the "exact"
background. In other words, the tachyon wave functions are given in
terms of functions over $Sl(2,R)$ \refs\rDVV , and the harmonic
analysis on the group should provide the analogs of the modular
covariant Dedekind functions $\eta(\tau)$ as building blocks of $\bar
\Lambda$.

 The other case is Rindler space, very interesting from the physical
point of view because all horizons look locally like Rindler's. Since
Rindler space is a reparametrization of Minkowski space, the
appropriate Conformal Field Theory should follow by  field
redefinitions from some free field formulation. In particular the
point particle world-line action \wlaction\ has the form,
$$
S_R [x^{\mu}(\tau)] = {1\over 4}\int \left\{ \left({d\theta\over
d\tau}\right)^2 + \left({dx\over d\tau}\right)^2 + 4\left[
\left({dx_{\perp}\over d\tau}\right)^2 +m^2 \right] e^{4\pi T_H
x}\right\}
$$
i.e. exponential barrier quantum mechanics, which is related to the
free case by B{\"a}cklund transformation . For the string case
the optical sigma model is of Liouville type and it is known that the
B{\"a}cklund transformation sends the theory onto a free one.
This machinery might be enough to produce a modular invariant one loop
free energy for strings in Rindler space starting from the analog
model expression as a sum of fields. Such modular invariant formulas
are important in order to generalize the answer to higher orders in
string perturbation theory.

 Yet another possibility is to follow the work of de Vega and Sanchez
\refs\rdeve ,  and start
from the light-cone quantization of strings in Rindler space (in flat
space the light-cone computation of the free energy leads directly to
the analog model \refs\real). Interestingly enough, these
authors make use of the horizon cutoff in order to regulate the
stretching effect of the string as it approaches the horizon.

 In any case, independently of the particular way in which these
technicalities are worked out, the physics of any string ensemble
based on the analog model seems clear.
The horizon divergences of each term would add up for the
string free energy, at least until the Hagedorn temperature is reached
locally. What happens beyond $T_{Hag}$ is presently a matter of
personal taste. If we adopt a Kosterlitz-Thouless scenario for the
phase transition \refs\rkog ,
          the compact coordinate of length $\beta$
is disordered by the vortices and all $\beta-$dependence would
disappear (see \refs\rgrokleb ). Under these
            circumstances $T_{Hag}$ acts as
a maximum temperature because nothing similar to "temperature" seems
to remain. In this scenario strings are probably regular at the
horizon but it is difficult to state anything more precise. This
scenario has the additional problem that it applies to two-dimensional
strings which do not really have Hagedorn transition (they are
essentially two-dimensional field theory plus the sporadic ``discrete"
states).

 On the other hand, in the Atick-Witten scenario \refs\raw , a first order
phase transition is triggered by a tachyonic winding mode. It is
argued that in the new phase the free energy scales (to all orders in
perturbation theory) as
\eqn\aw
{F(T) \sim T^2 }
i.e. like in a two-dimensional field theory. Hence, according to this
situation as the local temperature raises above $T_{Hag}$ we swich to
the law \aw\ and we get logarithmic horizon divergences in any
dimension. It is interesting to remark that \aw\ is compatible with
the extension by duality, which again holds in our case thanks to the
decoupling of the ``time" field in the optical background.

\newsec{Discussion}

 In this paper we have emphasized the infrared interpretation of
horizon divergences in Field Theory. We think that the correct
interpretation of these infinities as ultraviolet problems involves
considering the full dynamics, including the gravitational one. In
perturbative String Theory the space-time background is fixed and the
physics of the divergences seems similar as long as the Hagedorn
transition is not reached. Beyond this point the appropriate
phase of String
Theory is not known and the result depends on the particular scenario
we hold. For example, in the Atick-Witten view we would still obtain
horizon divergences, although of logarithmic type in any dimension.

 However, there are reasons to believe that String perturbation theory
alone is not enough to handle these divergences. In fact, it seems
that violent processes take place at the horizon once we switch the
gravitational back-reaction. First of all, it is well known that the
Canonical Ensemble is ill-defined in the presence of gravity because
of the Jeans instability, according to which self-gravitating
matter/energy becomes unstable under long wavelength fluctuations for
large enough volume. Nevertheless, a pragmatic attitude would be to
consider small enough volumes to be free from the Jeans collapse, but
still large enough  to be thermodynamical. But this condition is not
met in the vicinity of the horizon, due to the infinite blue-shift of
local temperatures. The Jeans length is estimated as:
$$
l_J \sim {1\over (G\rho)^{1/2}} \sim {\lambda_{W}^{2} \over l_P}
(\sqrt{-g_{00}})^{3/2}
$$
where  $\lambda_W \sim \hbar c / k_B T$ and $l_P$ is the Planck
length. If we adjust  room temperature far away from the horizon
we have
$$
l_J \sim 10^{25} \left( {\Delta s \over R_s}\right)^{3/2}\,\,
{\rm cm} $$
where $\Delta s$ is the (small) proper distance to the horizon and
$R_s$ stands for the Schwarzschild radius. So we see that
for super-massive black holes we can arrange the Jeans lenth at the
horizon to be $\sim$1 cm!. Of course the Hawking temperature of such a
black hole would be very small and this is the statement that violent
black hole condensation takes place at the boundary between two phases
with very different temperatures. If the asymptotic temperature is
taken to be the Hawking temperature then the Jeans mechanism occurs at
planckian distances from the horizon.

 On top of these effects one could also add the nucleation of Planck
size black holes discussed by Gross et. al. \refs\rgro ,
 which also becomes unsuppressed
as the temperature approaches Planck value. The main lesson here is
that independently of how massive is our black hole (how flat it is at
the horizon) in certain physical situations like thermal equilibrium,
the back reaction becomes completely uncontrollable at the horizon. The
type of processes involved seem to be rather non-perturbative and this
makes unlikely that they are described by String perturbation theory.

 It is very amusing to analytically continue these answers back to
real time in the context of the information paradox. Here the analog
of holding finite temperature at infinity is the constraint of
measuring finite energy {\it out} states at late times, and the analog of
the divergent local temperatures is given by the divergent frequencies
needed at intermediate stages in the semiclassical analysis. So, if we
admit that the requirement of thermal equilibrium makes the horizon
highly unstable even in the semiclassical limit, we would suspect that
$S$-matrix questions make no sense as long as we keep the standard
Penrose diagram for the background. This is strongly reminiscent of
recent ideas in \refs\rthof , who formulate S-matrix
problems in Penrose diagrams without horizons.
 In any case, it seems clear that the instabilities of gravitational
thermodynamics are closely related to the information paradox, and
this makes the study of two-dimensional almost-solvable models doubly
interesting.

\newsec{Acknowledgements}

I am indebted to K. Demeterfi and H. Verlinde for discussions. This
work is supported by NSF PHY90- 21984 grant.

\newsec{Note Added}

 The infrared problem and the subtleties involved in the
computation of determinants (section 3) were also noticed
in ref. \refs\rgidd , in the context of Schwinger pair production
near horizons.

\listrefs
\bye